\documentclass[a4paper]{article}

\usepackage{amsmath}
\usepackage[dvips]{graphicx}
\usepackage{hangcaption}

\title{Dynamics of gas sphere under self-gravity}

\author{Souichi Murata\footnote{Corresponding author.E-mail address: smurata@allegro.phys.nagoya-u.ac.jp}, \   Kazuhiro Nozaki \\\\
Department of Physics, Nagoya University, Chikusa-ku, Nagoya, 464-8602, Japan
}

\begin{document}

\maketitle
\begin{abstract}
\ \ \ \ A new dynamical solution for a gas sphere under self-gravity is 
presented to describe  a development of a gas sphere  from a motion-less state
to a state of expansion with a constant speed and a reflection
phenomenon in the dynamics of the surface of the sphere. \\

\end{abstract}

\section{Introduction}

\ \ \ \ The polytropic model of stellar structure is often introduced in studies of the different stages of stellar evolution. The equation governing the static equilibrium of a polytropic gas sphere is called the Endem equation, whose solution representing the structure of star is known as the Endem solution \cite{chand}.
  Quasi-statical change of stellar structure expressed by the Endem solution is believed to describe the evolution of star when the Endem solution is dynamically stable. The dynamical stability is not guaranteed unless the polytropic index $N$ is less than 3 \cite{hayashi}. A gas sphere of $N=3$ is borderline to the dynamical stability and is also interesting since the polytropic index $3$ corresponding to the adiabatic constant $\gamma=4/3$. There are some interesting stellar gases with $\gamma=4/3$, such as a radiation pressure dominant gas, an 
extremely relativistic gas and a gas with degenerate electrons \cite{hayashi}. 

  In order to extend the static Endem solution to dynamical states,      
Munier and Feix derived an explicit self-similar solution which describes a 
decelerating expansion of a gas sphere
of $N=3$ \cite{munier}.  Recently, another explicit self-similar solution has 
been obtained to describe a non accelerating nor decelerating expansion (or contraction) of a gas sphere of $N=3$ \cite{munier}.

In this paper,we present a new dynamical solution for a gas sphere of $N=3$,
which  is not self similar in general but includes the three known Endem-type solutions described above as special cases. The new dynamical solution describes such interesting phenomena as a development of a gas sphere  from a motion-less state to a state of expansion with a constant speed and a reflection phenomenon in the dynamics of the surface of a sphere.

\section{Dynamical solution of gas sphere}

\ \ \ \ The equations governing the spherically symmetrical flow of a 
polytropic gas of adiabatic index $\gamma$  under the influence of its 
own gravitation  are

\begin{align}
\rho_t+u\rho_r+\rho \bigg(u_r+\frac{2}{r}u \bigg)&=0, \label{cont}\\
u_t+u u_r+\rho^{\gamma-2} \rho_r+\sigma_r&=0,  \label{motion}\\
\sigma_{rr}+\frac{2}{r}\sigma_r&=\rho, \label{Poisson}
\end{align}
where the pressure $P$, density $\rho$, radial velocity $u$, 
gravitational potential $\sigma$  and radius $r$  are normalized by 
$P_0$, $\rho_0$, $u_0$ = $(\gamma P_0/\rho_0)^{1/2}$, $\gamma 
P_0/\rho_0$, $r_0$, respectively; $G$ is the gravitational constant and 
$P=\rho^{\gamma}$. The acoustic time scale $r_0/u_0$ , 
where $r_0 = \gamma P_0/(4\pi G \rho_0^2)$, is set to equal to the time 
scale of free fall $(4 \pi G \rho_0)^{-1/2}$. The system of Eqs. (\ref{cont}), (\ref{motion}) and  
(\ref{Poisson})
  admits a following Lie point symmetry

\begin{align}
\mathbf{V} \equiv 
(t+t_0)\partial_t-(\gamma-2)r\partial_r-2\rho\partial_{\rho}
                    -(\gamma-1)u\partial_{u}-\gamma\phi\partial_{\phi},
                                                         \label{symmetry}
\end{align}
where $\phi = \sigma_r$ and $t_0$ is an arbitrary constant. The 
symmetry (\ref{symmetry}) is derived through the standard procedure. Solving the Lie equation associated with the infinitesimal generator 
(\ref{symmetry}),  we  introduce   a new independent variable $T$ 
defined as

\begin{align}
dT=\frac{dt}{t+t_o},  \label{dT}
\end{align}

and scaled variables

\begin{align}
\rho&=\frac{R(x,T)}{(t+t_0)^2},\quad 
u=\frac{U(x,T)}{(t+t_0)^{\gamma-1}},
                               \quad 
\phi=\frac{\Phi(x,T)}{(t+t_0)^{\gamma}},
                                                        \label{invariant}
\intertext{where}
x&=\frac{r}{t^{2-\gamma}},\quad T=\log (t+t_0).\nonumber
\end{align}

Let us make the following ansatz

\begin{align}
R(x,T)&=\tilde{R}(y)
            \exp \bigg\{ \int \bigg( 2-3\overline{U} \bigg) dT 
\bigg\},
                                              \label{ansatz_R}\\
U(x,T)&=\overline{U}(T)x,\quad               \label{ansatz_U} \\
\Phi(x,T)&=\tilde{\Phi}(y)
            \exp \bigg\{  \int \bigg( \gamma-2\overline{U}\bigg) 
dT\bigg\},
                                              \label{ansatz_Phi}
\intertext{where}
y&=x\exp\bigg\{\int \bigg(2-\gamma-\overline{U} \bigg)dT \bigg\}. \nonumber
\end{align}
This ansatz ensures that  Eq. (\ref{cont}) is satisfied automatically.
For $\gamma=4/3$ , Eqs. (\ref{motion}) and (\ref{Poisson}) lead to

\begin{align}
ay+\tilde{R}^{\gamma-2}\tilde{R}_y+\tilde{\Phi}&=0, \label{remotion} \\
\overline{U}_T+\overline{U}(\overline{U}-1)
          &=a\exp \bigg\{\int\bigg(2-3\overline{U}\bigg)dT \bigg\},
                                            \label{U_relation} \\
\tilde{\Phi}_y+\frac{n}{y}\tilde{\Phi}&=\tilde{R}, \label{rePoisson}
\end{align}
where $a$ is an arbitrary constant.
Eqs. (\ref{remotion}) and (\ref{rePoisson}) give a modified Endem 
equation for  the density profile $\tilde{R}(y)$

\begin{align}
\theta_{y'y'}+\frac{2}{y'}\theta_{y'}+\theta^3+3a=0, \quad
                 \theta = \tilde{R}^{1/3}, \quad y'= \frac{y}{\sqrt{3}}.
                                             \label{theta}
\end{align}
According to numerical integration of Eq.(\ref{theta}) with the 
boundary condition $\theta(0)=1,\quad \theta_{y'}(0)=0$, the density profile has the first zero at finite $y'= z$ for $a_0 >-0.00219$ as shown in  Figure. 1. 
The value of $z$ determines the radius of  a gas sphere.      For $a=0$, $z$  takes the Endem's static value $z= 6.89685$.  Since the term $ay'$  comes from the inertia term and represents acceleration effect for $a>0$,
%%%%%%%%%%%
   we call $a$ as the acceleration parameter in this paper.
  %%%%%%%%%%%
In the acceleration case, the surface of  a gas sphere feels  inward inertia force and shrinks from the  Endem's static value.  For $a<0$, the surface 
feels outward inertia force and expands. At $a\approx-0.00219$, $z$ 
diverges to infinity and a localized gas sphere disappears.

\begin{figure}[t]
\rotatebox{-90}{
\scalebox{0.3}{
\includegraphics{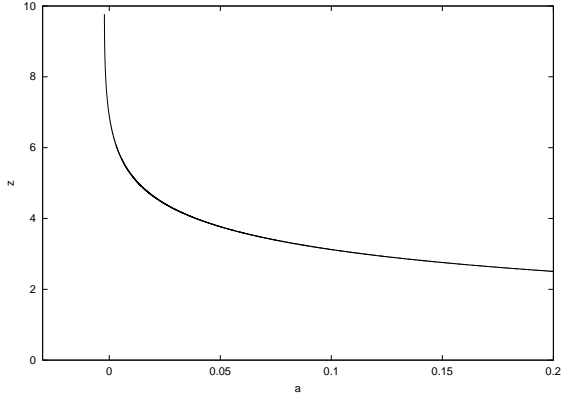}
}
}
\caption{z vs. the acceleration parameter $a$}
\end{figure}

It is easy to see that Eq.(\ref{U_relation}) takes the following 
integrable form
\begin{align}
f_{tt}=\frac{a}{f^2}, \label{U_relation1}
\end{align}
where
\begin{align}
f(t)=\exp\bigg(\int^T UdT\bigg).
\end{align}
Eq. (\ref{U_relation1}) can be integrated once and yields

\begin{align}
\frac{1}{2}f_t^2+\frac{a}{f}=b, \label{f_relation}
\end{align}
where $b$ is an integration constant. In terms of $f(t)$,
the solution  (\ref{invariant}) is rewritten as
\begin{align}
y=\frac{r}{f},\quad \rho=\frac{\tilde{R}(y)}{f^{n+1}},\quad
u=f_t y, \quad \phi=\frac{\tilde{\Phi}(y)}{f^n},\label{solution}
\end{align}
  where $\tilde{R}(y)$ is expressed by a localized solution of 
Eq.(\ref{theta})
while Eq.(\ref{remotion}) gives $\tilde{\Phi}(y)$ in terms of $\tilde{R}(y)$.
It is confirmed that the solution  (\ref{solution}) satisfies the 
boundary condition at the moving surface of sphere. While the surface 
is moving with a speed $\tilde{u}=\sqrt{3} f_t z$, the flow speed at the 
surface is given by the same one, i.e. $u= f_t y=\sqrt{3}f_t z$.\\
When  $b=0$ and $a<0$, Eq.(\ref{f_relation}) yields

\begin{align}
f(t)=\bigg( \frac{3\sqrt{-2a}}{2}(t-t_0) \bigg)^{2/3}   \nonumber
\end{align}
 and the solution (\ref{solution}) becomes identical with the known 
self-similar solution \cite{munier}.  In the absence of the inertia term ( 
i.e. $a=0$),  Eq.(\ref{theta}) reduces to the Endem equation. 
For $b=0$ and $a=0$, $f(t)$ is constant and Eq.(\ref{solution}) gives the static Endem solution. For $b>0$ and $a=0$,  Eq.(\ref{f_relation}) gives $f(t)=\sqrt{b}t$   and  Eq.(\ref{solution})  reproduces the expanding Endem solution which is expanding with the constant acoustic speed \cite{smurata}.
For the other values of $a$ and $b$, a new class of expanding solution 
with positive or negative acceleration is obtained.
Since $\tilde{u}=\sqrt{3}f_t z$, the acceleration rate of the surface is given  by $\tilde{u}_t=a\sqrt{3}z/f^2$,
of which sign is the same as the $a$'s sign. 

To demonstrate the evolution of the solution (\ref{solution}) , we 
consider an initial value problem for Eq.(\ref{f_relation}) such that 
  $f(t)|_{t=0}=1$ in order that $y|_{t=0}=r$. 

First of all, let us consider the case in which $a > 0$ and the 
acceleration rate
of the surface is radially positive. Since $f   \to  \infty$  as $t \to 
\infty$,
  $f_t  \to \sqrt{2b}$ asymptotically and  the surface eventually 
expands with a constant speed $\sqrt{6b}z$.  In this case, we can take a 
motionless initial condition so that  $f_t|_{t=0}=0$ or $u|_{t=0}=0$ by 
choosing the arbitrary (positive) constant $b$ as $b=a(>0)$.  The 
expansion speed of the surface ($\tilde{u}=\sqrt{3}z f_t$) is 
depicted for $a=b=0.01$ in Fig.2.

\begin{figure}[t]
\rotatebox{-90}{
\scalebox{0.3}{
\includegraphics{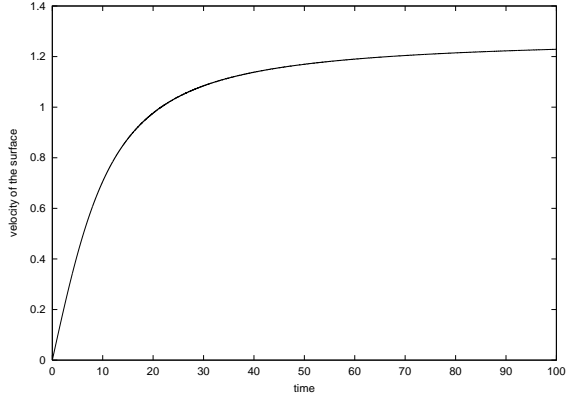}
}
}
\caption{Expansion velocity of the surface  for a=b=0.01.}
\end{figure}

Next, we consider the radially decelerating case (i.e. $a < 0$ ), where
we  take  $f_t|_{t=0}>0 $ as an initial condition so that the initial 
velocity of
the surface is positive.  Then, the initially expanding surface 
gradually decelerates
due to the negative inertia term.  In this case, Eq.(\ref{f_relation}) reads

\begin{align}
\frac{1}{2}f_t^2=b+\frac{|a|}{f}.
\end{align}
For $b> 0$, there are no reflection points and the expanding speed of 
the surface
decreases to a constant speed , that is,

\begin{align}
\lim_{t\to\infty}\tilde{u}=\sqrt{6b}z, \quad
\lim_{t\to\infty}\tilde{u}_t=0.
\end{align}
We illustrate the velocity of  the surface for $a=-0.002$ and  $b=0.001$  in Fig. 3. \\
When  $b<0$,  a reflection point appears at  $f=a/b$.  The deceleration 
of the surface stops at the reflection point and then the gas sphere 
began to shrinks.
The reflection time $t_r$ is given as
When $a < 0$ and  $b < 0$, the equation (\ref{f_relation}) yields

\begin{align}
t_r=\bigg|\frac{\sqrt{(2b-2a)}}{2b}
   -\frac{a}{b\sqrt{-2b}}\arctan\sqrt{ \frac{2b-2a}{-2b} }
\bigg|.   \nonumber
\end{align}

In this reflection case, the velocity of the surface is depicted
for $a=-0.002$ and $b=-0.001$ in Fig.4, where $t_r=57.485$.

\begin{figure}[t]
\rotatebox{-90}{
\scalebox{0.3}{
\includegraphics{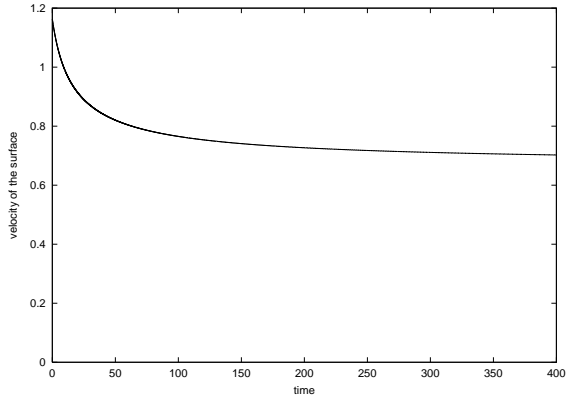}
}
}
\caption{Velocity of  the surface for a=-0.002 and b=0.001.}
\end{figure}

\begin{figure}[t]
\rotatebox{-90}{
\scalebox{0.3}{
\includegraphics{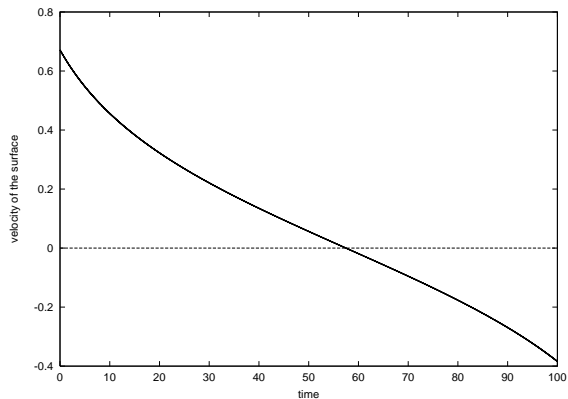}
}
}
\caption{Velocity of  the surface for a=-0.002 and b=-0.001.}
\end{figure}

\section{summary}

We present a new dynamical solution for a gas sphere under 
self-gravity, which not only unifies the three Endem-type solutions for  the polytropic index  $N=3$
but also describes the following interesting phenomena. \\
For positive values of acceleration parameter $a$, the radius of the 
gas sphere
decreases from the Endem's radius due to inward inertia force, while
the gas sphere expands for the negative $a$ and the distinct surface 
disappears
when $a<-0.00219$.
The new solution also describes acceleration  of the surface of a gas 
sphere  from a motion-less state to a state of expansion with a 
constant speed for the positive
$a$ and a reflection
phenomenon in the dynamics of the surface of a sphere for the negative 
$a$.

\end{document}